\documentclass[prb,twocolumn,showpacs,amsmath,amssymb,floatfix]{revtex4-1}
\usepackage{graphicx}
\usepackage{dcolumn}
\newcolumntype{d}[1]{D{i}{i}{#1}}
\usepackage{longtable}

\usepackage{epstopdf}

\begin{document}

\title{Modulated, three-directional, and polar structural instability
  in layered $d^1$ NaTiO$_2$}

\author{Alaska Subedi} 

\affiliation{Centre de Physique Theorique, Ecole Polytechnique, CNRS,
  Universit\'e Paris-Saclay, F-91128 Palaiseau, France}
\affiliation{Coll\`ege de France, 11 place Marcelin Berthelot, 75005
  Paris, France}

\date{\today}

\begin{abstract}
  I study the experimentally observed metal-to-metal structural phase
  transition in NaTiO$_2$ using density functional calculations. I do
  not find the previously proposed low-temperature structure
  energetically favorable with respect to the high-temperature
  rhombohedral structure. The calculated phonon dispersions of the
  rhombohedral phase show dynamical instabilities at several
  inequivalent parts of the Brillouin zone, including at the
  wavevector $(\frac{1}{2},\frac{1}{5},\frac{1}{5})$. These
  instabilities lead to monoclinic structures without inversion
  symmetry that are modulated along all three directions. The
  calculated electronic structures show that a local bonding
  instability of the Ti $3d$ states is associated with the structural
  transition.
\end{abstract}


\maketitle

\section{Introduction}

Layered NaTiO$_2$ has garnered attention because of its potential as
an anode material in sodium-ion batteries.\cite{mazz83,wu15} In
addition, this material also exhibits a broad metal-to-metal
structural transition between 200 and 250 K that has puzzled
researchers for more than twenty years.\cite{take92,clar96,clar98}
Unusual structural transitions are at the heart of several complex
phase diagrams, including that of the manganites that show colossal
magnetoresistance and the cuprate and iron unconventional
superconductors. Unraveling the mechanism of the structural transition
in NaTiO$_2$ and understanding its low temperature phase may
illustrate the range of possibilities for structural instabilities in
partially filled $3d$ systems. This may also help us understand the
causes of structural failures of battery materials, which plague the
performance of solid-state batteries.

Above 250 K, NaTiO$_2$ occurs in a high-symmetry layered rhombohedral
structure with space group $R\overline{3}m$.\cite{hage62} This
$\alpha$-NaFeO$_2$-type structure is a modification of the rock-salt
structure, and it consists of positive Na$^{+}$/Ti$^{3+}$ and negative
O$^{2-}$ ions at the sites of two interpenetrating fcc lattices, with
the NaO$_6$ and TiO$_6$ octahedra alternating along the rock-salt
[111] direction. Within each layer, the octahedra share edges so that
Na$^{+}$/Ti$^{3+}$ ions make equilateral triangles. In the
out-of-plane direction, the NaO$_6$ octahedra are elongated with
respect to the in-plane distances, while the TiO$_6$ octahedra are
compressed. As a consequence, the Na$^{+}$ and Ti$^{3+}$ ions are in
antiprismatic ($\overline{3}m$), rather than octahedral
($m\overline{3}m$), coordination.

NaTiO$_3$ formally contains $d^1$ Ti$^{3+}$ ions with spin $S =
\frac{1}{2}$ arranged in a triangular lattice. The initial interest in
this material came from the possibility of realizing a resonating
valence bond state of spin singlet pairs as suggested by
Anderson.\cite{ande73,faze74,hira85,yama85} Experimentally, however,
this material is metallic and shows a small, almost temperature
independent magnetic susceptibility above 250
K.\cite{take92,clar96,clar98} As the temperature is lowered,
signatures of a broad phase transition are seen in neutron
diffraction, magnetic susceptibility, and heat capacity
experiments. Several diffraction peaks of the high-temperature
rhombohedral structure split during the phase transition, and the
low-temperature structure has been resolved by Clarke \textit{et
  al.}\ to be in space group $C2/m$ with shortened Ti-Ti
bonds.\cite{clar96,clar98} This continuous phase transition is not
accompanied by a smooth change in the Ti-Ti bond lengths. Instead, the
rhombohedral and monoclinic phases with different bond lengths coexist
over the temperature range of the transition, and the fraction of the
monoclinic phase progressively increases as the temperature is
lowered. The magnetic susceptibility decreases continuously between
250 and 200 K, remains almost constant until 100 K, and then starts to
rise again, presumably due to impurities, as the temperature is
further lowered. Heat capacity measurements show a peak near 260 K,
which provides an unmistakable evidence for a bulk phase
transition. None of the aforementioned measurements exhibit any
hysteresis behavior.

The metal-to-metal transition in NaTiO$_2$ has been described in terms
of increased bandwidth due to stronger Ti-Ti bonds.\cite{clar98} This
picture relies on an electronic instability that lifts the orbital
degeneracy of the $d^1$ Ti$^{3+}$ electron. Several theoretical
studies have studied the electronic properties of NaTiO$_2$ and tried
to identify a mechanism for such an
instability,\cite{ezho98,pen97,khom05,dhar12,dhar14} but the
microscopic basis for the phase transition has still not been fully
clarified.

In this paper, I reexamine the metal-to-metal structural phase
transition observed in NaTiO$_2$ using density functional
calculations. I do not find the experimentally proposed
low-temperature structure energetically favorable compared to the
high-temperature rhombohedral structure. However, the calculated
phonon dispersions of the rhombohedral phase do show dynamical
instabilities at several inequivalent places in the Brillouin
zone. The structural distortions corresponding to the unstable phonon
modes cause modulations along both in-plane and out-of-plane
directions and lead to a monoclinic phase without inversion
symmetry. The wavevectors of the unstable phonons do not nest the
Fermi sheets. Hence, the monoclinic phase cannot be characterized as a
charge-density-wave state. Instead, I find that the transition is due
to a local bonding instability of the partially filled Ti $3d$ states.

\section{Computational Details}

The results presented here were obtained from calculations based on
density functional theory within the local density approximation. The
structural and magnetic relaxations were performed using the
pseudopotential-based planewave method as implemented in the Quantum
{\sc espresso} package. The phonon dispersions were calculated using
density functional perturbation theory.\cite{dfpt} The cut-offs were
set to 50 and 500 Ryd for the basis-set and charge-density expansions,
respectively. The pseudopotentials generated by Garrity \textit{et
  al.}\ were used in these calculations.\cite{gbrv} I used a $14
\times 14 \times 14$ grid for Brillouin zone sampling, and equivalent
or denser grids were used in the supercell calculations. The dynamical
matrices were calculated on an $8 \times 8 \times 8$ grid, and the
phonon dispersions were obtained by Fourier interpolation. All the
results presented in this paper were calculated using the fully
relaxed structures in the respective phases. Some calculations were
also performed using the general full-potential linearized augmented
planewave method as implemented in the {\sc wien2k}
package.\cite{wien2k}

The phonon frequencies except the unstable branch are converged to
less than a percent with the above cut-offs and grids, and a
Marzari-Vanderbilt smearing of 0.02 Ryd.\cite{mv} The unstable branch
was not converged even when using a computationally-demanding
$k$-point grid of $18\times18\times18$, and I let it be because the
unstable frequencies cannot be measured anyway. I confirmed the
presence of the structural instability by stabilizing supercell
structures that are distorted according to the unstable phonons. A
drawback for not converging the unstable branch is that I am not fully
sure of the wavevector that has the strongest instability.

\section{Results and Discussions}

Several theoretical studies based on both first principles and model
calculations have attempted to clarify the instability in NaTiO$_2$
that leads to the low-temperature $C2/m$ structure proposed by Clarke
\textit{et al.}\cite{ezho98,pen97,khom05,dhar12,dhar14} Instead of
trying to search for the mechanism that leads to the proposed
low-temperature phase, I started my investigation by trying to
stabilize the experimentally determined structure. However, when I
fully relaxed the experimental $C2/m$ structure, it relaxed to the
high-temperature rhombohedral structure with $R\overline{3}m$
symmetry. This was true regardless of whether I allowed a magnetic
solution or not. Even if I started with a magnetically ordered state
with large initial moments, the system relaxed to the rhombohedral
structure with no or very small moments. This lack of a strong
magnetic instability is consistent with the experiments, which show
almost temperature independent magnetic susceptibility with a small
value of $\chi \approx 4.5 \times 10^{-4}$ emu/mol for the
high-temperature phase.\cite{take92,clar98}

\begin{figure}
  \includegraphics[width=\columnwidth]{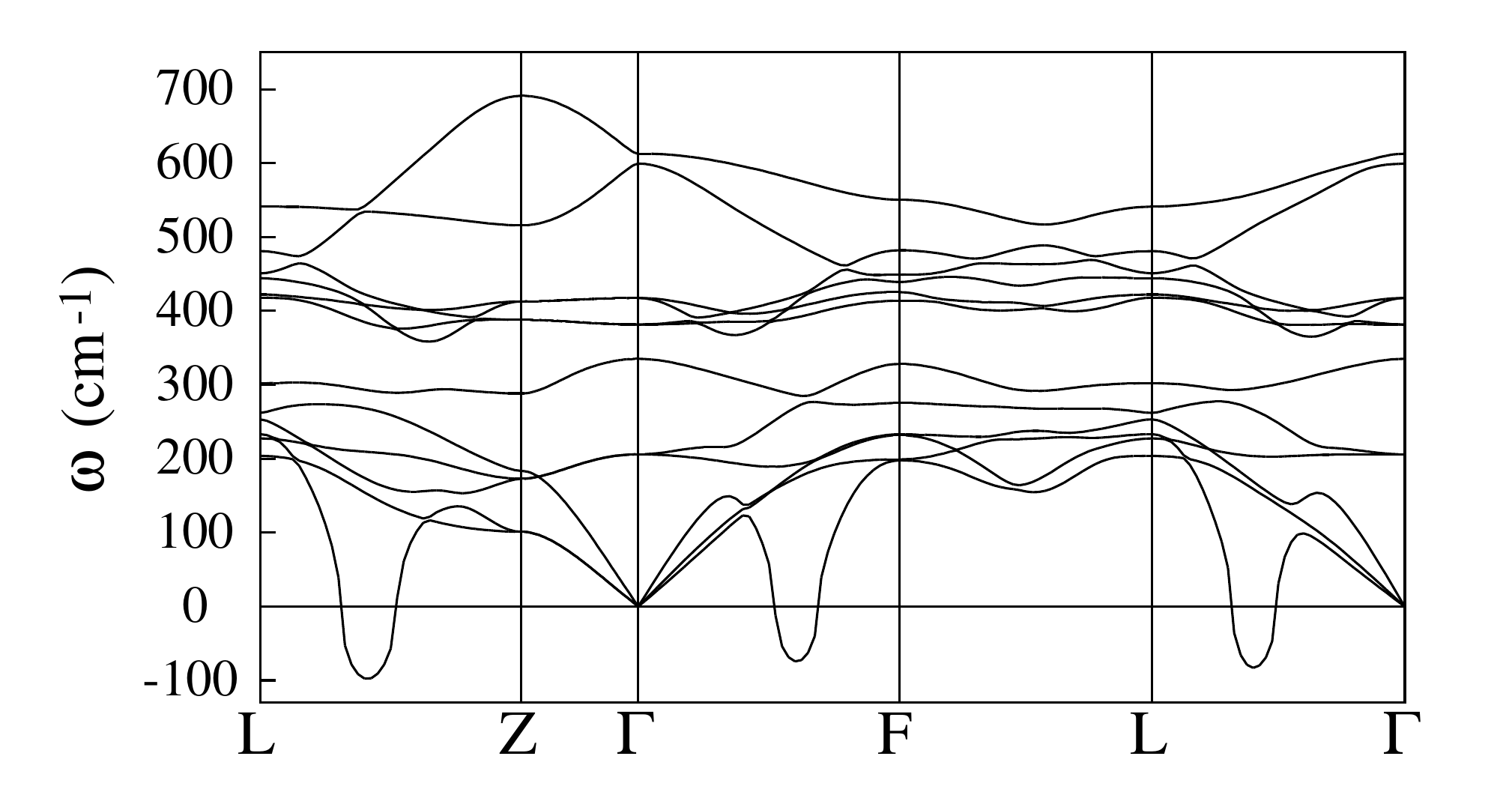}
  \caption{Calculated phonon dispersions of NaTiO$_2$ in the
    rhombohedral structure with space group $R\overline{3}m$. The
    dispersions are plotted along the path $L\ (\frac{1}{2}, 0, 0)
    \rightarrow Z\ (\frac{1}{2}, \frac{1}{2}, \frac{1}{2}) \rightarrow
    \Gamma\ (0,0,0) \rightarrow F\ (\frac{1}{2}, \frac{1}{2}, 0)
    \rightarrow L\ (\frac{1}{2}, 0, 0) \rightarrow \Gamma\ (0, 0, 0)$
    in the Brillouin zone of the primitive unit cell. The imaginary
    frequencies are represented with negative values.}
  \label{fig:pband}
\end{figure}

I calculated the phonon dispersions of rhombohedral NaTiO$_2$ in the
theoretical structure to examine if this material has any dynamical
instabilities as suggested by the experiments.  The relaxed
rhombohedral unit cell has the hexagonal lattice parameters of $a =
2.9235$ and $c = 16.2167$ \AA\ and the internal O parameter of $z =
0.2370$.  The phonon dispersions, shown in Fig.~\ref{fig:pband}, do
not display any instabilities at or near the Brillouin zone center. In
particular, the acoustic branches do not exhibit any softness even
along the Cartesian $z$ direction $\Gamma$$\rightarrow$$Z$
(\textit{i.e.} the rhombohedral [111] direction). In fact, the
acoustic branch that is polarized along the out-of-plane direction has
the largest dispersion. This demonstrates the presence of strong
interlayer bonding despite the layered structure of this material.


\begin{table}
  \caption{\label{tab:eigdisp} The eigendisplacement vector of the
    unstable phonon mode of rhombohedral NaTiO$_2$ at the wavevector
    $(\frac{1}{2}, \frac{1}{5}, \frac{1}{5})$.}
  \begin{ruledtabular}
     \begin{tabular}{l d{12.0} d{12.0} d{12.0}}
       atom & \multicolumn{1}{c}{$x$} & \multicolumn{1}{c}{$y$} &
       \multicolumn{1}{c}{$z$} \\
       \hline 
       Na & -0.024 + 0.012 i &  0.042 - 0.022 i &  0.112 - 0.057 i \\
       Ti &  0.248 - 0.250 i & -0.430 + 0.432 i & -0.062 + 0.063 i \\
       O  &  0.037 + 0.052 i & -0.065 - 0.090 i &  0.002 + 0.472 i \\
       O  & -0.052 - 0.037 i &  0.090 + 0.064 i & -0.472 + 0.000 i 
    \end{tabular}
  \end{ruledtabular}
\end{table}

A conspicuous feature in the phonon dispersions is the presence of a
phonon branch that is unstable along the $L$--$Z$, $\Gamma$--$F$ and
$L$--$\Gamma$ paths. I find that the largest instability is at the
wavevector $(\frac{1}{2}, \frac{1}{5}, \frac{1}{5})$ in terms of the
primitive reciprocal lattice vectors of the rhombohedral cell. The
atomic displacements corresponding to this wavevector occur along all
three directions, and it is remarkable that a three-directional
structural instability could occur for a layered material.  The
unstable phonon mode has an irreducible representation $A'$ of the
point group $m$, which is a subgroup of the point group
$\overline{3}m$ of the high-temperature structure. This implies that
the distortions corresponding to the unstable mode leads to a
structure with a point group symmetry $m$. This is a crystal class
without inversion symmetry. The eigendisplacement vector of this
unstable mode is given in Table~\ref{tab:eigdisp}. As one can see,
this mode causes displacements of all four atoms of the rhombohedral
unit cell. The loss of inversion symmetry is also evident from the
asymmetric displacement pattern of the cations relative to that of the
negatively charged O$^{2-}$.

I performed full structural relaxations of 200-atom $2 \times 5 \times
5$ supercells to be certain that the structural instability is
real. When the starting structure consisted of a supercell with
randomly displaced atomic positions of the high-temperature structure,
the relaxation converged back to the high-symmetry rhombohedral
structure. However, when I started with a supercell that had atoms
displaced according to the displacement pattern obtained from the
eigendisplacement vector of the unstable mode, the relaxation yielded
a lower-symmetry structure whose total energy was only $\sim$1.4
meV/atom lower than that of the rhombohedral structure. The changes in
bond lengths due to the structural distortions are relatively
small. The Na-O and Ti-O bond lengths change by between 0.001 to 0.135
\AA\ (corresponding to changes of 0.05--6.0\%), which encompasses the
range of 0.5--1.5\% changes in bond lengths experimentally determined
by Clarke \textit{et al.} However, the distortion pattern of the
NaO$_6$ and TiO$_6$ octahedra is complex, owing to the wavevector
$(\frac{1}{2}, \frac{1}{5}, \frac{1}{5})$ corresponding to the
instability that repeats the distortions over a comparatively large
distance.  The low-symmetry structure can be refined to a ten formula
unit (40 atoms) monoclinc unit cell with space group $Cm$, and the
full structural information is given in Table~\ref{tab:str255} in the
Appendix. Note that this structure lacks inversion symmetry. However,
the off-centerings of the cations in the oxygen cages are small, and
the structure can also be refined to $C2/m$ symmetry if the tolerance
in the symmetry finder code is increased to 0.03 \AA. Therefore, the
results of my structural relaxations are not completely inconsistent
with Clarke \textit{et al.}'s experiment that resolves the
low-temperature structure to the space group $C2/m$. The volume of my
low-symmetry structure is same as that of the high-symmetry structure,
again in agreement with the experiment.

Similar dynamical instabilities occur at other points in the Brillouin
zone of the rhombohedral unit cell, and I also performed structural
relaxations of a 12-atom $1 \times 1 \times 3$ supercell. Again, the
supercell relaxed to a low-symmetry $Cm$ structure when the starting
structure consisted of distortions according to the eigendisplacement
vector of the unstable mode at the wavevector $(0,0,\frac{1}{3})$. The
total energy of this low-symmetry structure is $\sim$0.9 meV/atom
lower than that of the high-symmetry rhombohedral structure. The total
energies of the low-symmetry structures due to dynamical instabilities
at wavevectors $(\frac{1}{2}, \frac{1}{5}, \frac{1}{5})$ and
$(0,0,\frac{1}{3})$ are so close, and the distortions so similar (both
phonons belong to the $A'$ irreducible representation of the point
group $m$), that the distortions due to these instabilities could
coexist below the transition temperature. It is worth emphasizing that
the experimentally observed phase transition in NaTiO$_2$ is
continuous.  The picture of a structural phase transition caused by
finite-wavevector phonon instabilities that lead to low-symmetry
structures that are almost degenerate in energy is consistent with
this experimental evidence.

\begin{figure}
  \includegraphics[width=0.9\columnwidth]{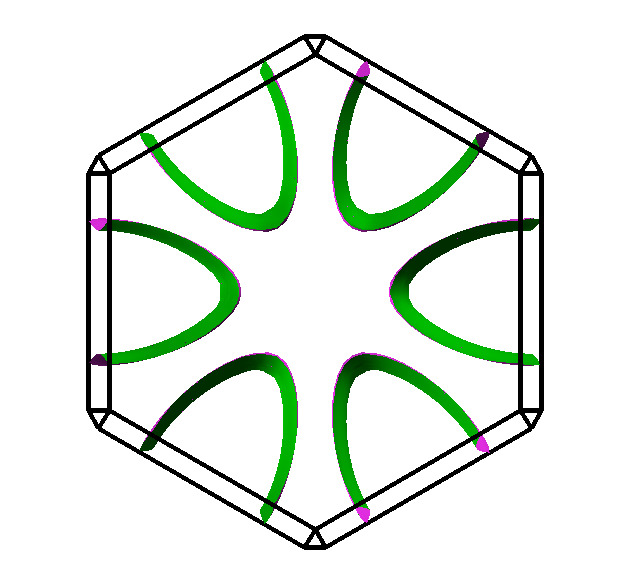}
  \caption{(Color online) Calculated Fermi surface of rhombohedral
    NaTiO$_2$.}
  \label{fig:fs}
\end{figure}

What might be the cause of this phase transition? The usual suspect
for the cause of a finite-wavevector structural distortion is charge
density wave instability due to a Fermi surface nesting. I studied the
electronic structure of rhombohedral NaTiO$_2$ to investigate the
possibility of this instability. The calculated Fermi surface of this
material is shown in Fig.~2, and it consists of fairly two-dimensional
elliptical cylinders that cross the six faces of the Brillouin zone
boundary. Since the Fermi sheets are elliptical, there are no
pronounced nestings. Furthermore, the stable phonon branches do not
show any Kohn anomalies at the wavevectors corresponding to the
imaginary phonon frequencies. Therefore, a charge density wave
instability cannot be the cause of the structural phase transition
experimentally observed in this material.

\begin{figure}
  \includegraphics[width=\columnwidth]{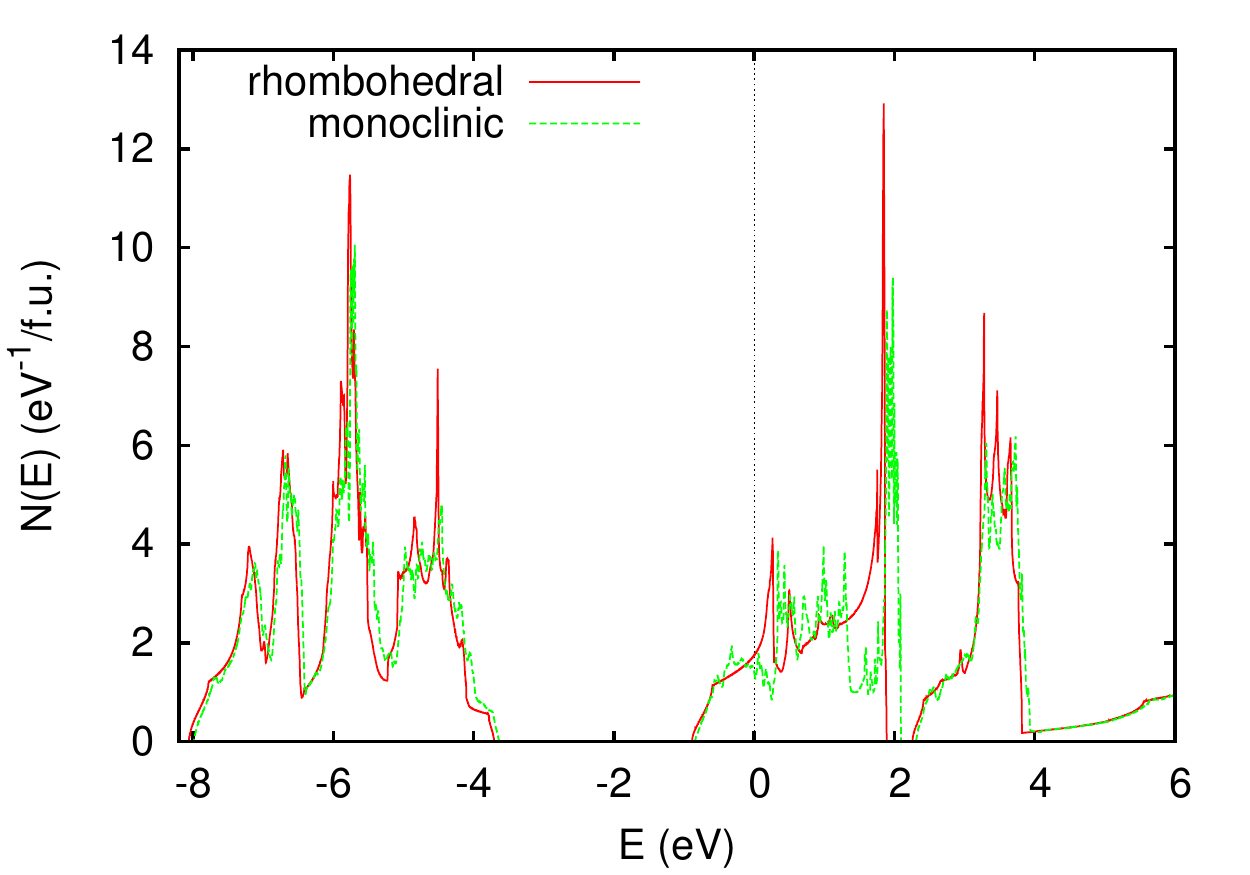}
  \includegraphics[width=\columnwidth]{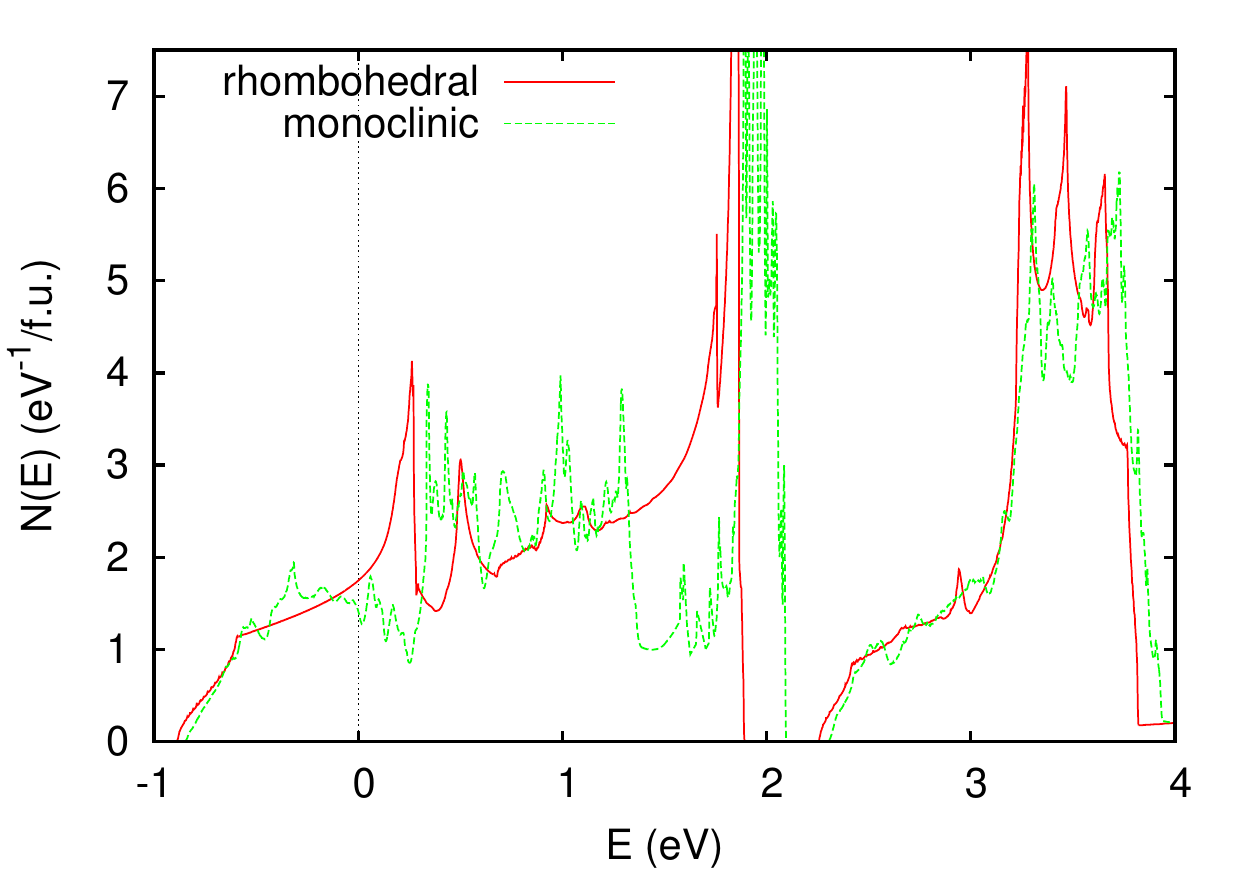}
  \caption{(Color online) Calculated electronic density of states of
    rhombohedral and monoclinic NaTiO$_2$. The monoclinic structure
    refers to the distortion due to the
    $(\frac{1}{2},\frac{1}{5},\frac{1}{5})$ phonon instability. The
    lower panel is a blowup of the Ti $3d$ manifold. }
  \label{fig:dos}
\end{figure}

\begin{figure}
  \includegraphics[width=\columnwidth]{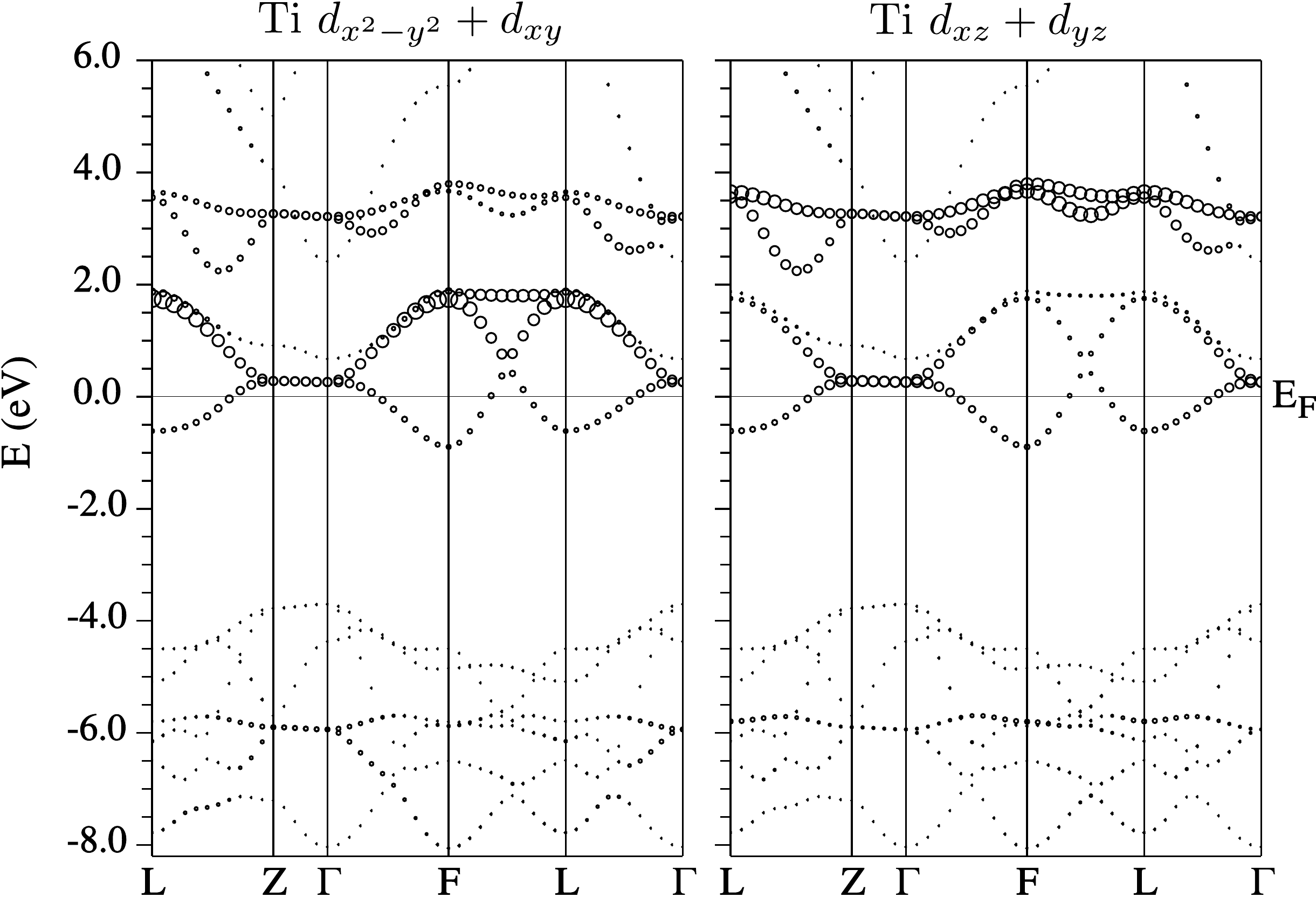}
  \caption{ Calculated band structures of rhombohedral NaTiO$_2$
    plotted with circles of size proportional to the Ti $d_{x^2-y^2} +
    d_{xy}$ (left) and Ti $d_{xz} + d_{yz}$ (right) characters. The
    orbitals are defined in the Cartesian coordinates of the
    conventional unit cell.}
  \label{fig:eband}
\end{figure}

A comparison of the electronic structures of rhombohedral and
monoclinic NaTiO$_2$ also gives further evidence that the structural
transition is not related to the instability of the Fermi surface. The
calculated electronic density of states (DOS) of the two phases are
shown in Fig.~3, and it shows that electronic states over a wide range
of energy are modified. This is in contrast to a charge density wave
instability, which would only change electronic states near the Fermi
energy. I do find that the DOS at the Fermi level is reduced from 1.75
eV$^{-1}$ per formula unit in the rhombohedral phase to 1.41 eV$^{-1}$
per formula unit in the monoclinic cell due to the
$(\frac{1}{2},\frac{1}{5},\frac{1}{5})$ phonon instability, a
reduction of 19\%. However, unlike in a charge density wave
transition, where one would observe an increase in the density of
states immediately above the charge density wave--induced gap, there
is actually a large decrease and redistribution of DOS immediately
above the Fermi energy.


The electronic states between $-$8.10 and $-$3.60 eV have dominant O
$2p$ character. As the band structure of the rhombohedral phase
plotted in Fig.~\ref{fig:eband} shows, these states also exhibit some
Ti $3d$ character, and this reflects the formally bonding O $2p$-Ti
$3d$ nature of these states. The states between $-$0.90 and 3.85 eV
show mostly Ti $3d$ character, and these are formally antibonding. The
Fermi energy corresponds to this manifold being occupied by one
electron.  In the point group $\overline{3}m$ of the rhombohedral
phase, the five Ti $3d$ orbitals can be grouped into states with
irreducible representations $e_g$ and $a_{1g}$. There are two sets of
$e_g$ states in this point group, namely $(d_{x^2-y^2}, d_{xy})$ and
$(d_{xz}, d_{yz})$. Nominally, the upper Ti $3d$ manifold consisting
of two bands between 2.25 and 3.85 eV would correspond to the $d_{xz}$
and $d_{yz}$ states, and the lower two bands in the lower manifold
would be $d_{x^2-y^2}$ and $d_{xy}$ states.  However, there is
significant mixing of orbital characters between the two manifolds due
to hopping---either direct or via O $2p$ orbitals---between
neighboring Ti ions. In addition, the nominally $a_{1g}$ band with
dominant $d_{z^2}$ character that lies in the upper part of the lower
Ti $3d$ manifold also shows $d_{x^2-y^2}$ and $d_{xy}$ characters, and
the lower $e_g$ manifold correspondingly shows a noticeable $d_{z^2}$
character. This mixing of $3d$ orbitals in this structure type has
also been emphasized in the context of $R\overline{3}m$
NaCoO$_2$.\cite{joha04,land06,land08}.  Therefore, the narrow $e_g$
manifold crossing the Fermi level should not be viewed as being
strongly correlated due to the effect of a large on-site Coulomb
repulsion $U$. The effective $U$ for the orbitals corresponding to
this manifold should be strongly reduced because these orbitals would
be quite extended with large tails not only at O sites, but also at
neighboring Ti ions when other manifolds are integrated out. However,
this is not to say that the calculated bandwidth will not be
renormalized, and it would be interesting to perform beyond density
functional theory calculations to study the effects of correlations.

Coming back to Fig.~3, it is striking to note that small differences
in the atomic positions between the rhombohedral and monoclinic phases
cause a large reconstruction of the electronic structure. The peak at
0.26 eV above the Fermi energy in the rhombohedral phase, which is
formed by states with mixed $d_{xz}, d_{yz}, d_{x^2-y^2},$ and
$d_{xy}$ characters, collapses in the monoclinic phase. Similarly, the
peak in the DOS near the upper edge of the lower $e_g$ manifold that
is made out of states with $d_{z^2}$ character moves higher in energy
in the monoclinic phase. These peaks are caused by the narrow
dispersion of the bands in the out-of-plane direction, and it reflects
the layered nature of this material. The large changes in the shapes
and positions of these peaks establish that the Ti $3d$ states and
changes in the effective hybridization in the out-of-plane direction
play a key role in the structural transition.

The shape of the DOS curve changes greatly due to the structural
transition, but there is in fact little change in the bandwidths of
the O $2p$ or Ti $3d$ manifolds. The bandwidth of the Ti $3d$ manifold
increases by 0.1 eV, which is small compared to its overall bandwidth
of 4.75 eV. The gap between the two $e_g$ manifolds, which is 0.38 eV
in the rhombohedral phase, shows a larger change and decreases to 0.22
eV in the monoclinic phases. This again indicates the importance of
effective Ti-Ti interaction for the structural instability. The
comparatively small role of Ti-O bonding in the transition is
supported by the relatively minor changes in the shape and position of
the O $2p$ manifold. The Na electronic states, which lie above 4 eV
relative to the Fermi energy, show almost no change. This indicates
that Na ions play an insignificant role in the structural instability.

Although NaTiO$_2$ is a rare example of a system that exhibits a
finite-wavevector modulated structural distortions due to a local (in
the sense of pertaining to immediate neighbors) bonding instability,
it is not the only material that shows such an instability. The
transition to a modulated structure observed in IrTe$_2$ has also been
described in terms of a local bonding instability associated with the
Te $5p$ states.\cite{fang13,cao13} NaTiO$_2$ illustrates that such a
local instability can also occur in transition metal oxides with
partially filled $3d$ orbitals.

\section{Summary and Conclusions}

In conclusion, I have found that modulated structural distortions
occur in $d^1$ NaTiO$_2$. These distortions occur in all three
directions and are different from those previously reported for the
low-temperature experimental structure. The structural distortions
correspond to unstable phonon modes that are nominally polar in
nature. The unstable phonon modes occur at several inequivalent parts
of the Brillouin zone. I relaxed 200-atom and 12-atom supercells
corresponding to the instabilities at the wavevectors $(\frac{1}{2},
\frac{1}{5}, \frac{1}{5})$ and $(0,0,\frac{1}{3})$, respectively, and
found that noncentrosymmetric monoclinic structures with space
group $Cm$ are stabilized. These structures are lower in energy by
only 1.4 and 0.9 meV/atom, respectively, than the high-temperature
rhombohedral phase. Such a small energy difference is consistent with
the continuous nature of the phase transition observed in this
material. The phonon instabilities do not correspond to any Fermi
surface nestings, and the structural instability cannot be described
as being of the charge density wave type. The distortions are due to a
local bonding instability of the Ti $3d$ states, which changes the
electronic structure over a wide range of energy. Similar bonding
instability associated with the Te $5p$ states has been discussed in
IrTe$_2$. The results discussed here show that such an instability can
also occur in a partially filled $3d$ system.

\acknowledgments 

I am grateful to Gwen\"aelle Rousse for helpful discussions.  This
work was supported by the European Research Council grants ERC-319286
QMAC and ERC-61719 CORRELMAT and the Swiss National Supercomputing
Center (CSCS) under project s575.

\appendix

\section{Structure}

The full structural details of the structures corresponding to the
phonon instabilities at the wavevectors $(\frac{1}{2},\frac{1}{5},
\frac{1}{5})$ and $(0,0,\frac{1}{3})$ are given in
Tables~\ref{tab:str255} and \ref{tab:str003}, respectively. These were
obtained by first fully relaxing the respective $2 \times 5 \times 5$
and $1 \times 1 \times 3$ supercells, and then refining thus obtained
structure using the {\sc findsym} code.\cite{findsym}

\newcolumntype{d}[1]{D{.}{.}{#1}}
\begin{table}[t]
  \caption{\label{tab:str255} Atomic positions of the fully-relaxed
    monoclinic structure with $Cm$ space group corresponding to the
    unstable phonon mode at
    $(\frac{1}{2},\frac{1}{5},\frac{1}{5})$. The lattice parameters
    are $a = 24.4313$, $b = 2.9241$, $c = 11.3368$ \AA, $\alpha =
    90^\circ$, $\beta = 98.8352^\circ$, and $\gamma = 90^\circ$. }
  \begin{ruledtabular}
     \begin{tabular}{l d{1.5} d{1.5} d{1.5}}
       atom & \multicolumn{1}{c}{$x$} & \multicolumn{1}{c}{$y$} &
       \multicolumn{1}{c}{$z$} \\
       \hline 
        Na  &    0.00049  &   0.00000  &  -0.00117 \\
        Na  &    0.40066  &   0.00000  &   0.39881 \\
        Na  &    0.79993  &   0.00000  &   0.80004 \\
        Na  &    0.19929  &   0.00000  &   0.20135 \\
        Na  &    0.59962  &   0.00000  &   0.60096 \\
        Na  &   -0.00048  &   0.00000  &   0.50067 \\
        Na  &    0.39931  &   0.00000  &  -0.09847 \\
        Na  &    0.80007  &   0.00000  &   0.29980 \\
        Na  &    0.20073  &   0.00000  &   0.69843 \\
        Na  &    0.60037  &   0.00000  &   0.09957 \\
        Ti  &    0.89817  &   0.00000  &   0.64883 \\
        Ti  &    0.29624  &   0.00000  &   0.04763 \\
        Ti  &    0.69924  &   0.00000  &   0.44956 \\
        Ti  &    0.10305  &   0.00000  &   0.85188 \\
        Ti  &    0.50314  &   0.00000  &   0.25193 \\
        Ti  &   -0.09862  &   0.00000  &   0.15079 \\
        Ti  &    0.30385  &   0.00000  &   0.55238 \\
        Ti  &    0.70107  &   0.00000  &  -0.04930 \\
        Ti  &    0.09650  &   0.00000  &   0.34788 \\
        Ti  &    0.49735  &   0.00000  &   0.74842 \\
         O  &   -0.04771  &   0.00000  &   0.30514 \\
         O  &    0.35249  &   0.00000  &   0.71218 \\
         O  &    0.75242  &   0.00000  &   0.11283 \\
         O  &    0.15297  &   0.00000  &   0.50731 \\
         O  &    0.55329  &   0.00000  &  -0.09689 \\
         O  &   -0.04739  &   0.00000  &   0.81022 \\
         O  &    0.35326  &   0.00000  &   0.20453 \\
         O  &    0.75284  &   0.00000  &   0.60314 \\
         O  &    0.15223  &   0.00000  &   0.00856 \\
         O  &    0.55246  &   0.00000  &   0.41363 \\
         O  &    0.84698  &   0.00000  &  -0.00296 \\
         O  &    0.24681  &   0.00000  &   0.39472 \\
         O  &    0.64749  &   0.00000  &   0.78899 \\
         O  &    0.04751  &   0.00000  &   0.18660 \\
         O  &    0.44781  &   0.00000  &   0.59258 \\
         O  &    0.84758  &   0.00000  &   0.48666 \\
         O  &    0.24756  &   0.00000  &   0.88865 \\
         O  &    0.64758  &   0.00000  &   0.29564 \\
         O  &    0.04669  &   0.00000  &   0.69667 \\
         O  &    0.44714  &   0.00000  &   0.09180 \\
    \end{tabular}
  \end{ruledtabular}
\end{table}

\begin{table}[b]
  \caption{\label{tab:str003} Atomic positions of the fully-relaxed
    monoclinic structure with $Cm$ space group corresponding to the
    unstable phonon mode at $(0,0,\frac{1}{3})$. The lattice
    parameters are $a = 10.9416$, $b = 2.9232$, $c = 8.6562$ \AA,
    $\alpha = 90^\circ$, $\beta = 119.8852^\circ$, and $\gamma =
    98.8352^\circ$. }
  \begin{ruledtabular}
     \begin{tabular}{l d{1.5} d{1.5} d{1.5}}
       atom & \multicolumn{1}{c}{$x$} & \multicolumn{1}{c}{$y$} &
       \multicolumn{1}{c}{$z$} \\
       \hline 
       Na  &   0.00026  &   0.00000  &  -0.00067 \\
       Na  &   0.33324  &   0.00000  &   0.66695 \\
       Na  &   0.66649  &   0.00000  &   0.33364 \\
       Ti  &   0.66982  &   0.00000  &   0.83891 \\
       Ti  &  -0.00500  &   0.00000  &   0.49110 \\
       Ti  &   0.33523  &   0.00000  &   0.17002 \\
        O  &   0.31661  &   0.00000  &  -0.08064 \\
        O  &   0.64591  &   0.00000  &   0.58974 \\
        O  &  -0.01538  &   0.00000  &   0.25265 \\
        O  &   0.02076  &   0.00000  &   0.74493 \\
        O  &   0.34856  &   0.00000  &   0.41536 \\
        O  &   0.68350  &   0.00000  &   0.07800 \\
    \end{tabular}
  \end{ruledtabular}
\end{table}

\end{document}